\documentclass[12pt]{iopart}
\usepackage{iopams}
\usepackage{url,graphicx}

\begin{document}

\title{Highly entangled multi-qubit states with simple algebraic structure}

\author{J E Tapiador$^{1}$, J C Hernandez-Castro$^{2}$, J A Clark$^{1}$ and S Stepney$^{1}$}

\address{$^1$ Department of Computer Science, University of York, UK}
\address{$^2$ Department of Computing, University of Portsmouth, UK}

\ead{jet@cs.york.ac.uk}

\begin{abstract}
Recent works by Brown \etal \cite{Brown05} and Borras \etal \cite{Borras07}
have explored numerical optimisation procedures to search for highly
entangled multi-qubit states according to some computationally tractable
entanglement measure. We present an alternative scheme based upon
the idea of searching for states having not only high entanglement but
also simple algebraic structure. We report results for 4, 5, 6, 7 and 8 qubits
discovered by this approach, showing that \emph{many} of such states do exist.
In particular, we find a maximally entangled 6-qubit state with an algebraic structure
simpler than the best results known so far. For the case of 7 qubits, we discover
states with high, but not maximum, entanglement and simple structure, as well
as other desirable properties. Some preliminary results are shown for the case
of 8 qubits.
\end{abstract}

\submitto{\JPA}

\maketitle

\section{Introduction}\label{Sec:Intro}
The concept of entanglement is intimately related to the mathematical
structure of quantum mechanics, particularly as a direct consequence of
linearity in tensor product Hilbert spaces. Apart from its theoretical
relevance as one (if not the most) striking features of quantum mechanics,
entanglement has been shown to enable (and play a crucial role in) practical
applications such as teleportation protocols and superdense coding.
It is, for this reason, often regarded as a \emph{resource} which needs
to be studied from several standpoints.

A considerable amount of research has been devoted to unveil the mathematical
structures underlying entanglement, in particular concerning its
\emph{quantification}. Properties for a good entanglement measure are
reviewed and discussed in \cite{Horo00, Horo01, PV06}. An alternative to
the analytic approach was presented by Brown \etal \cite{Brown05} and later
by Borras \etal \cite{Borras07}. Both works explore the application of
numerical optimisation techniques to search for highly entangled multi-qubit
states.

In the work of Brown \etal \cite{Brown05}, the search is applied over the set of mixed
states using the negativity of the partial transpose as a measure of
entanglement. The search procedure is basically a hill-climbing algorithm
with some adjustments. Such a simple form of search suffices, as the cost
function which is sought to maximise (i.e. the entanglement as defined
by the negativity) is known to be convex. Among the results reported are
the following two 5-qubit states
\begin{equation}\label{BrownState5A}
\begin{array}{rl}
|\Psi^{''}\rangle = & \frac{1}{2\sqrt{2}}\big(
|00110\rangle + |01011\rangle + |10001\rangle + |11100\rangle\\
& + i(|00101\rangle + |01000\rangle + |10010\rangle + |11111\rangle)
\big)
\end{array}
\end{equation}
\begin{equation}\label{BrownState5B}
\begin{array}{rl}
|\Psi^{'''}\rangle = & \frac{1}{2\sqrt{2}}\big(
|00110\rangle + |01001\rangle + |10101\rangle + |11010\rangle\\
& + i(|00000\rangle + |01111\rangle + |10011\rangle + |11100\rangle)
\big)
\end{array}
\end{equation}
which are essentially equivalent and can be reformulated as
\begin{equation}\label{BrownState5}
|\psi_5\rangle = \frac{1}{2}(
|001\rangle|\Phi_{-}\rangle + |010\rangle|\Psi_{-}\rangle +
|100\rangle|\Phi_{+}\rangle + |111\rangle|\Psi_{+}\rangle)
\end{equation}
where $|\Psi_{\pm}\rangle = \frac{1}{\sqrt{2}}|00\rangle \pm |11\rangle$ and
$|\Phi_{\pm}\rangle = \frac{1}{\sqrt{2}}|01\rangle \pm |10\rangle$. The state
$|\psi_5\rangle$ achieves the maximum possible entanglement
according to the negativity measure. Muralidharan and Panigrahi
\cite{MP08a, MP08b} have recently investigated the usefulness of this
state for quantum teleportation, superdense coding and quantum
secret sharing. In particular, perfect teleportation of
arbitrary 1- and 2-qubit systems is possible, as well as
superdense coding of 5 classical bits by using three qubits.
(Incidentally, it is also shown how to physically produce
this state in a relatively easy manner.)

In the work of Borras \etal \cite{Borras07}, the search is restricted
to pure states of 4, 5, 6 and 7 qubits, and several measures of entanglement
are investigated. The overall search algorithm is very similar to the
hill-climbing scheme employed by Brown \etal, and the authors also explore
the statistical distribution of entanglement for systems of the sizes
aforementioned. In the case of 6 qubits, they discovered the state
\begin{equation}\label{BorrasState6}
\begin{array}{rl}
|\psi_6\rangle = \frac{1}{\sqrt{32}}( &
|000000\rangle + |111111\rangle + |000011\rangle + |111100\rangle + |000101\rangle +\\
& |111010\rangle + |000110\rangle + |111001\rangle + |001001\rangle + |110110\rangle +\\
& |001111\rangle + |110000\rangle + |010001\rangle + |101110\rangle + |010010\rangle +\\
& |101101\rangle + |011000\rangle + |100111\rangle + |011101\rangle + |100010\rangle -\\
& |001010\rangle + |110101\rangle + |001100\rangle + |110011\rangle + |010100\rangle +\\
& |101011\rangle + |010111\rangle + |101000\rangle + |011011\rangle + |100100\rangle +\\
& |011110\rangle + |100001\rangle)\\
\end{array}
\end{equation}
which, as in the case of Brown \etal's $|\psi_5\rangle$, possesses a rather
simple structure and maximum entanglement according to several measures.
Choudhury \etal \cite{CMP09} have also found various applications for this
state, namely quantum teleportation of an arbitrary 3-qubit state and
quantum state sharing of an arbitrary 2-qubit state.

In the case of 7 qubits, the authors report a highly entangled state
(approximately 155.812856, according to the negativity measure) with, apparently,
no simple structure. The state is given as a list of 128 complex
numbers corresponding to the coefficients of the superposition, and
has all the 1-qubit density matrices completely mixed, but not the
2- and 3-qubit ones.

\subsection{Motivation}
Apart from theoretical repercussion, genuinely entangled states of
5 and 6 qubits have found practical applications and new protocols
based on them have been derived. In the case of 7 qubits, it is not
known how to write ``nicely'' the state(s) reported by Borras \etal.
This lack of structure may be a drawback in deriving protocols based
on it.

The search schemes described above have proven to be very successful,
but the resulting states can be difficult to process manually. In this work,
we present an alternative formulation of the search where not only high
entanglement is sought, but also some desired structure is imposed on the
form of the outcome. As a proof-of-concept, we show how highly entangled
states with a very compact form can be easily found.

\section{Computationally feasible entanglement measures}\label{Sec:EntMeasures}
All the numerical optimisation approaches aforementioned rely directly
upon the use of a computationally tractable measure of entanglement.
For completeness and readability, we next provide a brief introduction
to some of them, and particularly to the measure used in this paper. For
a more in-depth treatment of the subject, we refer the reader to
\cite{Horo00, Horo01, PV06}.

The degree of \emph{bipartite} entanglement of a composite quantum system
can be quantified in terms of the purity (or mixedness) of the reduced
density matrix of one of the two subsystems: the lower the purity (or the
higher the mixedness), the higher the entanglement. There is not, however,
a unique way of extending this notion to systems composed of more than two
subsystems. A natural alternative is to quantify multipartite entanglement
among $n$ parties as a function of the mixedness of \emph{all}
the possible bipartitions of the whole system. A common, though not unique,
approach is to take the sum of the individual entanglement measures associated with
the $2^{n-1}-1$ possible bipartitions of a $n$-qubit system \cite{Brown05,Borras07}
\begin{equation}\label{Eq:MultipartiteEnt}
E(\rho) = \sum_{s} E_B(\rho_s)
\end{equation}
where $\rho$ is the density matrix of the $n$-qubit system, $\rho_s$ is the
reduced density matrix associated with subsystem $s$, and the sum is taken
over all the possible bipartions. Common examples of bipartite entanglement
measures $E_B$ are:
\begin{enumerate}
\item The linear entropy: $S_L(\rho) = 1 - \mathrm{Tr}(\rho^2)$
\item The von Neumann entropy: $S_{VN}(\rho) = - \mathrm{Tr}(\rho\log_2 \rho)$
\item The Renyi entropy with $q\rightarrow \infty$: $S_{Re}^{q\rightarrow \infty}(\rho) =
-\mathrm{ln} \lambda_k^{max}$, $\lambda_k$ being the eigenvalues of $\rho$.
\end{enumerate}

All these entropic measures constitute clear examples of computationally feasible
measures of entanglement for \emph{pure} states. They cannot, however, distinguish
between classical and quantum correlations, and so their use for mixed states is
quite limited.

Brown \etal developed in \cite{Brown05} an entanglement measure for mixed states
based on the negative partial transpose (NPT) criterion, which establishes as a
necessary condition for separability of any density matrix that it has only
non-negative eigenvalues \cite{Peres96}. Details about how to compute the partial
transpose can be found in \cite{Brown05}.

The \emph{negativity}, first introduced in \cite{Zycz98} and subsequently explored
in \cite{VW02,Kendon02,Brown05,Borras07}, is defined as the sum of all the negative
eigenvalues of the $2^{n-1}-1$ partial transposes. The result if often negated
in order to have a positive measure. We will refer to this measure as $E_{\mathrm{NPT}}$.
In some cases it is convenient to have a measure normalised between 0 and 1, which
can be achieved by dividing by the maximum possible entanglement. Upper bounds for
the four entanglement measures discussed above can be derived by considering a
hypothetical $n$-qubit pure state such that all its marginal density matrices are
completely mixed \cite{Borras07}. In the case of the $E_{\mathrm{NPT}}$, this
maximum is 1.5, 6.5, 17.5, 60.5, 157.5, 504.5, 1297.5, $\ldots$ for systems composed of
3, 4, 5, 6, 7, 8, 9, $\ldots$ qubits. We will denote by $E_{\mathrm{NPT}}^{\mathrm{norm}}$
the normalised entanglement.

\section{Searching for simplicity}\label{Sec:Search}

\subsection{Search space}
We want to search for pure states having high entanglement and \emph{simple} algebraic structure. There is not, however, a unique way to define what ``simple'' means, and some of such definitions might not be computationally appropriate. For the purposes of this work, simple states will be those which can be written in a ``nice'' way with respect to a given basis, i.e. states of the form
\begin{equation}\label{eq:state}
\displaystyle|\psi\rangle = \frac{1}{K}\sum_{i=0}^{2^n-1}c_i |i\rangle
\end{equation}
where only a few of the coefficients $c_i$ are non-null and, in turn, are nice to write (e.g. $\pm 1$, $\pm i$, $\pm (1\pm i)$, $\ldots$). Even though this is quite an arbitrary definition of simplicity, it captures well the intuitive idea of what a nice-to-write state means; and, more importantly, provides a \emph{measurable} way to determine how simple a state is, a crucial property if we are to somehow incorporate such an aspect into the search.

For the purposes of this paper, the space of ``simple'' states will be given by
\begin{equation}\label{eq:searchspace}
\mathbb{S}^n(V) = \bigg\{\displaystyle|\psi\rangle = \frac{1}{K}\sum_{i=0}^{2^n-1}c_i |i\rangle\quad : \quad c_i \in V\bigg\}
\end{equation}
where $1/K$ is the appropriate normalisation factor. The size and properties of such a space\footnote{Note that ``space'' is used here as in ``search space'' and not as in ``Hilbert space'', as $\mathbb{S}^n(V)$ might not always be one.} strongly depend on the set $V$ of allowed coefficients. Three examples that will be used later are $V_3=\{0, \pm 1\}$, $V_5=\{0, \pm 1, \pm i\}$ and $V_9=\{0, \pm 1, \pm i, \pm (1 \pm i)\}$.

Using a simple counting argument it is not difficult to see that the size of the search space is $|\mathbb{S}^n(V)| = |V|^{2^n}$. Note, however, that some states might actually be identical up to some global phase, and therefore previous expression is actually an upper bound on the number of useful states. The exact amount of indistinguishable states depends on the particular elements in the set $V$ and we will not be generally concerned about them. Table \ref{Table:SizeSearchSpace} shows the size of the search space for the number of qubits considered in this work and for some small values of $|V|$. The combinatorial explosion is easily recognisable in these figures, and any attempt of search by enumeration is beyond our current computational resources even for small values.

\begin{table}[t]
\caption{\label{Table:SizeSearchSpace}Upper bounds for $|\mathbb{S}^n(V)|$ for some values of $n$ and $|V|$.}
\begin{indented}
\item[]\begin{tabular}{cccccc}
\br
& \centre{5}{Number of qubits}\\
\ns
& \crule{5}\\
$|V|$ & 4 & 5 & 6 & 7 & 8\\
\mr
\multicolumn{1}{c}{5} & $2^{37.1}$ & $2^{74.3}$ & $2^{148.6}$ & $2^{297.2}$ & \multicolumn{1}{c}{$2^{594.4}$}\\
\multicolumn{1}{c}{7} & $2^{44.9}$ & $2^{89.8}$ & $2^{179.7}$ & $2^{359.3}$ & \multicolumn{1}{c}{$2^{718.7}$}\\
\multicolumn{1}{c}{9} & $2^{50.7}$ & $2^{101.4}$ & $2^{202.9}$ & $2^{405.7}$ & \multicolumn{1}{c}{$2^{811.5}$}\\
\multicolumn{1}{c}{11} & $2^{55.3}$ & $2^{110.7}$ & $2^{221.4}$ & $2^{442.8}$ & \multicolumn{1}{c}{$2^{885.6}$}\\
\multicolumn{1}{c}{13} & $2^{59.2}$ & $2^{118.4}$ & $2^{236.8}$ & $2^{473.6}$ & \multicolumn{1}{c}{$2^{947.3}$}\\
\br
\end{tabular}
\end{indented}
\end{table}

\subsection{Fitness function}
A good fitness (or cost) function should provide \emph{guidance} through the search space towards
the desired solutions, regardless of whether or not its numerical values are related
to the actual cost or benefit of particular solutions. As we are interested in
states having high entanglement and as many null coefficients as possible, a natural
choice for the fitness function is to explicitly reward both features. We have used
a fitness function of the form
\begin{equation}\label{eq:fitnessfunc}
F|\psi\rangle = E_{\mathrm{NPT}}^{\mathrm{norm}}|\psi\rangle\langle\psi| - \alpha
\bigg(\frac{N|\psi\rangle}{2^n}\bigg)
\end{equation}
where
\begin{equation}\label{eq:zerosfunc}
N(\sum_{i=0}^{2^n-1}c_i|i\rangle) = \#\big\{c_i : |c_i|\neq 0\big\}
\end{equation}
and $\alpha \in [0,1]$ is a punishing factor. The term $E_{\mathrm{NPT}}^{\mathrm{norm}}|\psi\rangle\langle\psi|$
simply computes the entanglement as measured by the NPT, while the right part counts the percentage of non-null coefficients in the superposition. Note that both terms are normalised in the interval $[0,1]$, so $\alpha$ admits a natural interpretation as how much punishment is placed on the amount on non-null coefficients. Other variations around the same idea are of course possible.

\subsection{Move function}
The move function determines which states are reachable from a given one
at any point of the search. We have defined a move function consisting of
randomly choosing one of the $2^n$ coefficients of the state, $c_i$, and
another, $c_j$, from $V$ such that $c_i\neq c_j$. The coefficient $c_i$
is replaced by $c_j$ and the state vector is then renormalised. Note that
each state vector has exactly $2^n(|V|-1)$ different\footnote{
As discussed above, some of them might be identical up to a global phase.}
neighbours. We will write $|\phi\rangle \leftarrow \mathrm{Move}|\psi\rangle$
to denote that state $|\phi\rangle$ is the result of moving state $|\psi\rangle$.

\subsection{Search procedure}
It is well known that the negativity cost function is convex \cite{VW02}
and its smoothness provides us with a relatively easy-to-walk search landscape.
In particular, its convexity for mixed states implies that there is no chance
for the search of being caught in a local optimum, even though the case for
pure states is less clear. However, the trade-off between entanglement and simplicity contained in expression
(\ref{eq:fitnessfunc}) translates into a more rugged search landscape,
with lower correlation between the fitness of neighbour state vectors and,
therefore, a substantial increase in the number and distribution of local optima.
This suggests that very simple forms of search as those used in previous works (basically
hill-climbing algorithms) are not likely to be appropriate in this case. We have verified
experimentally this fact and concluded that a more powerful search technique is necessary.

Simulated Annealing \cite{kirkpatrick83} is a well-known search heuristic
inspired by the cooling processes of molten metals. It
can be seen as a basic hill-climbing coupled with the
probabilistic acceptance of non-improving solutions. This
mechanism allows a local search that eventually can escape from
local optima.

The search starts at some initial state (solution) $|\psi\rangle \in_R \mathbb{S}^n$.
The algorithm employs a control parameter $T \in \mathbb{R}^+$ known
as the ``temperature''. This starts at some positive value $T_0$ and
is gradually lowered at each iteration, typically by geometric
cooling: $T_{i+1} = \beta T_i,~\beta \in [0,1]$.
At each temperature, a number MIL (Moves in Inner Loop) of
neighbour states are attempted. A candidate state $|\phi\rangle$ in the
neighbourhood of $|\psi\rangle$ is obtained by applying the move
function. The new state is always accepted if it is better than
$|\psi\rangle$. To escape from local optima, the
technique also accepts candidates which are slightly worse than
the current state, meaning that its fitness is no more than $|T \ln U|$ lower,
with $U$ a uniform random variable in $(0,1)$. As $T$ is gradually lowered,
this term gets closer to $0$ and it becomes harder to accept worse moves.

The algorithm terminates when some stopping criterion is met,
usually after executing a fixed number of inner loops or when
some maximum number of consecutive inner loops without improvements
have been reached. The basic algorithm is shown below.\\

\begin{footnotesize}
\begin{tabular}{rl}
1 & $|\psi\rangle \in_R \mathbb{S}^n$\\
2 & $|\psi^{max}\rangle \leftarrow |\psi\rangle$\\
3 & $T \leftarrow T_0$\\
4 & \textbf{repeat until} stopping criterion is met\\
5 & ~~~~~~~\textbf{repeat} MIL \textbf{times}\\
6 & ~~~~~~~~~~~~~~$|\phi\rangle \leftarrow \textrm{Move}|\psi\rangle$\\
7 & ~~~~~~~~~~~~~~$U \in_R (0,1)$\\
8 & ~~~~~~~~~~~~~~\textbf{if} $F|\phi\rangle > F|\psi\rangle + T \ln U$ \textbf{then}\\
9 & ~~~~~~~~~~~~~~~~~~~~~~~~$|\psi\rangle \leftarrow |\phi\rangle$\\
10& ~~~~~~~~~~~~~~~~~~~~~~~~\textbf{if} $F|\psi\rangle > F|\psi^{max}\rangle$ \textbf{then}\\
11& ~~~~~~~~~~~~~~~~~~~~~~~~~~~~~~~~~~$|\psi^{max}\rangle \leftarrow |\psi\rangle$\\
12& ~~~~~~~~~~~~~~~~~~~~~~~~\textbf{endif}\\
13& ~~~~~~~~~~~~~~\textbf{endif}\\
14& ~~~~~~~\textbf{endrepeat}\\
15& ~~~~~~~$T \leftarrow \beta T$\\
16& \textbf{endrepeat}\\
17& return $|\psi^{max}\rangle$\\
\end{tabular}
\end{footnotesize}

\subsection{Implementation}
The search scheme described above was implemented in C++, using the GNU Scientific
Library \cite{GSL} for complex numbers, matrix manipulations, pseudorandom
number generation, eigenvalues, etc.

\section{Results}\label{Sec:Results}
We next present and discuss some results obtained for systems of 4, 5, 6, 7 and 8 qubits.
A couple of states are shown for each case. These are representative of the
typical outcome of the search, but \emph{many} others have been produced by
the technique.

In all the cases, the search algorithm is parameterised with 1000 moves in the
internal loop (MIL) and a stop criterion consisting of 10 consecutive internal
loops without improvement. For the number of qubits studied, the best results
are achieved with $\alpha$ between 0.1 and 0.3, a starting temperature $T_0$
between 0.0005 and 0.001, and a cooling rate $\beta$ between 0.99 and 0.999.

\subsection{4-qubit states}
Higuchi and Sudbery proved in \cite{HS00} that there is no 4-qubit pure state having all its
marginal density matrices maximally mixed. Consequently, the theoretically maximum amount of
entanglement for systems of 4 qubits (e.g., 6.5 according to the negativity measure) is
unachievable. The state
\begin{equation}\label{state4HS}
|HS\rangle = \frac{1}{\sqrt{6}}\big(
|1100\rangle + |0011\rangle + \omega(|1001\rangle + |0110\rangle)
+ \omega^2(|1010\rangle + |0101\rangle)
\big)
\end{equation}
where $\omega=-\frac{1}{2} + \frac{\sqrt{3}}{2}i$ is also reported in \cite{HS00}. It is
conjectured to be maximally entangled, having
$E_{\mathrm{NPT}}|HS\rangle \approx 6.0981$. Borras \etal claim in \cite{Borras07} that their
search procedure consistently finds
states with exactly this same amount of entanglement. In fact, the HS state is known to be a
\emph{local} maximum
for the von Neumann entropy \cite{BH07}, and some arguments given in \cite{HS00} suggest
that it may also been a \emph{global} maximum for this measure.

In our experimentation using the set of coefficients $V_5$ we found the state
\begin{equation}\label{state4_a}
|\psi_4^a\rangle = \frac{1}{\sqrt{6}}\big(
|1001\rangle - i(|0000\rangle - |0011\rangle + |0110\rangle + |1100\rangle + |1111\rangle)\big)
\end{equation}
\begin{equation}\nonumber
E_{\mathrm{NPT}}|\psi_4^a\rangle \approx 5.989631
\end{equation}
This state has 22 negative eigenvalues from 7 partial transposes. For the four single-index cuts,
each partial transpose has a single negative eigenvalue of $-\frac{1}{2}$, and all the 1-qubit density
matrices are maximally mixed. The 18 two-index partial transposes have a more complex structure, each one
having 6 negative eigenvalues. None of the 2-qubit density matrices are maximally mixed.

When using the set $V_9$ we found the state
\begin{equation}\label{state4_b}
\begin{array}{rl}
|\psi_4^b\rangle = & \frac{1}{2\sqrt{6}}\big(
|0001\rangle - |0100\rangle + i(|1011\rangle - |1110\rangle) +\\
& (1+i)(|0010\rangle + |1101\rangle - |0101\rangle - |0110\rangle - |1010\rangle - |1100\rangle) +\\
& (1-i)(|1000\rangle + |1001\rangle - |0011\rangle - |0111\rangle)
\big)
\end{array}
\end{equation}
\begin{equation}\nonumber
E_{\mathrm{NPT}}|\psi_4^b\rangle \approx 6.051660
\end{equation}
As in the case of $|\psi_4^a\rangle$, this state has 22 negative eigenvalues from 7 partial transposes.
The four single-index cuts give a partial transpose with a single negative eigenvalue of $-\frac{1}{2}$, and all the 1-qubit density matrices are maximally mixed. The 18 two-index partial transposes have 6 negative
eigenvalues each one. Again, none of the 2-qubit density matrices are maximally mixed.

The HS state is obviously more entangled than $|\psi_4^a\rangle$ and $|\psi_4^b\rangle$, yet
both states (particularly the latter) do have a very high entanglement and a rather simple structure.
Merely for validation purposes, we also attempted searches including both $\omega$ and $\omega^2$
in the set of valid coefficients $V$. In such cases, the HS state (or states essentially equivalent to it)
is easily found.

\subsection{5-qubit states}
The search successfully finds 5-qubit states exhibiting maximum entanglement
($E_{\mathrm{NPT}}=17.5$) and a rather simple algebraic structure. The following
two, $|\psi_5^a\rangle$ and $|\psi_5^b\rangle$, are particular examples found
using coefficients in $V_5$ and $V_9$, respectively

\begin{equation}\label{state5_a}
\begin{array}{rl}
|\psi_5^a\rangle = & \frac{1}{2\sqrt{2}}\big(
|01010\rangle + |10011\rangle + |10110\rangle + |11000\rangle\\
& + i(|00001\rangle - |00100\rangle + |01111\rangle - |11101\rangle)
\big)
\end{array}
\end{equation}
\begin{equation}\nonumber
E_{\mathrm{NPT}}|\psi_5^a\rangle = 17.5
\end{equation}

\begin{equation}\label{state5_b}
\begin{array}{rl}
|\psi_5^b\rangle = & \frac{1}{4}\big(
(1+i)(|01101\rangle + |01110\rangle - |10100\rangle + |10111\rangle\\
& + |11000\rangle + |11011\rangle) + (1-i)(|00001\rangle - |00010\rangle)
\big)
\end{array}
\end{equation}
\begin{equation}\nonumber
E_{\mathrm{NPT}}|\psi_5^b\rangle = 17.5
\end{equation}

Both states have 65 negative eigenvalues from 15 partial transposes. For the
five single-index cuts, each partial transpose has a single negative eigenvalue
of $-\frac{1}{2}$. For the ten two-index cuts, each partial transpose has six
negative eigenvalues of $-\frac{1}{4}$. So
\begin{equation}\label{ent_psi5}
E_{\mathrm{NPT}}|\psi_5^a\rangle = E_{\mathrm{NPT}}|\psi_5^b\rangle = 5\cdot\frac{1}{2}
+ 10\cdot 6\cdot\frac{1}{4} = 17.5
\end{equation}

All the partial density matrices are completely mixed for both states, i.e.
$\mathrm{Tr}(\rho^2) = \frac{1}{2}$ for all the 1-qubit marginal density matrices,
and $\mathrm{Tr}(\rho^2) = \frac{1}{4}$ for all the 2-qubit marginal density matrices.

It is not difficult to show that these two states are equivalent under local unitary
transformations, and also equivalent to Brown \etal's $|\psi_5\rangle$ state given 
by (\ref{BrownState5}). All the maximally entangled 5-qubit states found in our
experimentation have this form.

\subsection{6-qubit states}
As with the case of 5 qubits, the search finds maximally entangled
($E_{\mathrm{NPT}}=60.5$) 6-qubit states having simple structure.
Two examples are provided below

\begin{equation}\label{state6_a}
\begin{array}{rl}
|\psi_6^a\rangle = & \frac{1}{4}\big(
- |000101\rangle - |001011\rangle - |010001\rangle + |011111\rangle\\
&+ |100000\rangle + |101110\rangle - |110100\rangle + |111010\rangle\\
&+ i(|000110\rangle - |001000\rangle - |010010\rangle - |011100\rangle\\
&+ |100011\rangle - |101101\rangle + |110111\rangle + |111001\rangle)
\big)
\end{array}
\end{equation}
\begin{equation}\nonumber
E_{\mathrm{NPT}}|\psi_6^a\rangle = 60.5
\end{equation}

\begin{equation}\label{state6_b}
\begin{array}{rl}
|\psi_6^b\rangle = & \frac{1}{4}\big(
|011010\rangle -|011111\rangle + |101011\rangle -|101110\rangle\\
& -|110011\rangle -|110110\rangle -|111000\rangle -|111101\rangle\\
& +i(-|000010\rangle -|000111\rangle + |001001\rangle + |001100\rangle\\
& +|010001\rangle -|010100\rangle -|100000\rangle +|100101\rangle)
\big)
\end{array}
\end{equation}
\begin{equation}\nonumber
E_{\mathrm{NPT}}|\psi_6^b\rangle = 60.5
\end{equation}

Both states have 376 negative eigenvalues from 31 partial transposes. For the
6 single-index cuts, each partial transpose has a single negative eigenvalue
of $-\frac{1}{2}$. For the 15 two-index cuts, each partial transpose has 6
negative eigenvalues of $-\frac{1}{4}$. For the 10 three-index cuts, each partial
transpose has 28 negative eigenvalues of $-\frac{1}{8}$. So
\begin{equation}\label{ent_psi6}
E_{\mathrm{NPT}}|\psi_6^a\rangle = E_{\mathrm{NPT}}|\psi_6^b\rangle = 6\cdot\frac{1}{2}
+ 15\cdot 6\cdot\frac{1}{4} + 10\cdot 28\cdot \frac{1}{8} = 60.5
\end{equation}

All the partial density matrices are completely mixed for both states, i.e.
$\mathrm{Tr}(\rho^2)$ being $\frac{1}{2}$, $\frac{1}{4}$ and $\frac{1}{8}$ for all the
1-qubit, 2-qubit and 3-qubit marginal density matrices, respectively.

\subsubsection{A new 6-qubit state}
The states given by (\ref{state6_a}) and (\ref{state6_b}) are equivalent under local unitary
transformations and can be reformulated as

\begin{equation}\label{Eq:State6Q_16coeffs}
|\Psi_6\rangle = \frac{1}{2}(
|F_0\rangle|\Psi_{-}\rangle + |F_1\rangle|\Psi_{+}\rangle +
|F_2\rangle|\Phi_{-}\rangle + |F_3\rangle|\Phi_{+}\rangle)
\end{equation}
with $|F_0\rangle = \frac{1}{\sqrt{2}}(|0000\rangle + |1111\rangle)$,
$|F_1\rangle = \frac{1}{\sqrt{2}}(|0011\rangle + |1100\rangle)$,
$|F_2\rangle = \frac{1}{\sqrt{2}}(|0110\rangle + |1001\rangle)$, and
$|F_3\rangle = \frac{1}{\sqrt{2}}(|0101\rangle + |1010\rangle)$. Note that
this state has an elegant description in terms of Bell pairs, since
\begin{equation}\label{Eq:State6Q_16coeffs}
\begin{array}{ll}
\frac{1}{\sqrt{2}}(|F_0\rangle + |F_1\rangle) = |\Psi_{+}\rangle|\Psi_{+}\rangle\\
\frac{1}{\sqrt{2}}(|F_2\rangle + |F_3\rangle) = |\Phi_{+}\rangle|\Phi_{+}\rangle
\end{array}
\end{equation}

This state has arguably a simpler algebraic structure than Borras \etal's $|\psi_6\rangle$ given
by (\ref{BorrasState6}). We postpone for future work a further study of the possible relationships
between both states.

\subsection{7-qubit states}
It has been pointed out that entanglement in 7-qubit systems might exhibit some
similarities with the case of 4 qubits. A conjecture by Borras \etal \cite{Borras07}
establishes that there is no pure state of 7 qubits whose marginal density matrices
for subsystems of 1, 2, or 3 qubits are all \emph{completely} mixed. If such
a state would exist, its entanglement as measured by the NPT would be 157.5. In
\cite{Borras07} it is reported that 7-qubit states with entanglement up to
$E_{\mathrm{NPT}}\approx 155.812856$ are found. They all have completely mixed
single-qubit marginal density matrices, but \emph{not} completely mixed 2- and
3-qubit marginal density matrices.

The results found with our approach are consistent with this numerical evidence.
The states $|\psi_7^a\rangle$ and $|\psi_7^b\rangle$ shown below have
high (but not maximum) entanglement. Contrarily to the 7-qubit state reported in
\cite{Borras07}, these states have a remarkably simple structure and, as will be
discussed later, totally mixed 1- and 2-qubit marginal density matrices.

The state

\begin{equation}\label{state7_a}
\begin{array}{rl}
|\psi_7^a\rangle = & \frac{1}{4\sqrt{2}}\big(
(1+i)(|0000010\rangle + |1000101\rangle + |1011011\rangle - |0011001\rangle\\
&- |1010000\rangle -|1111101\rangle) + (1-i)(|0001100\rangle + |0010111\rangle\\
&+ |0100001\rangle + |0101111\rangle + |0110100\rangle + |0111010\rangle + |1001110\rangle\\
&+ |1100011\rangle -|1101000\rangle -|1110110\rangle)\big)\\
\end{array}
\end{equation}
\begin{equation}\nonumber
E_{\mathrm{NPT}}|\psi_7^a\rangle = 152.646039
\end{equation}

\noindent have 1113 negative eigenvalues from 63 partial transposes. For the
7 single-index cuts, each partial transpose has a single negative eigenvalue
of $-\frac{1}{2}$. For the 21 two-index cuts, each partial transpose has 6
negative eigenvalues of $-\frac{1}{4}$. For 21 out of the 35 three-index cuts, each partial
transpose has 28 negative eigenvalues of $-\frac{1}{8}$; the remainder 14 partial
transposes have each 28 negative eigenvalues, but different from $-\frac{1}{8}$.

Both the 7 single-qubit and the 21 2-qubit marginal density matrices are all
completely mixed, i.e. with $\mathrm{Tr}(\rho^2)$ being $\frac{1}{2}$ and $\frac{1}{4}$,
respectively. For the case of the 35 3-qubit marginal density matrices, 21 of them
are completely mixed, corresponding to the following bipartitions
\begin{equation}\label{bipartitions7_aOK}
\begin{array}{l}
\{0,1,2\}, \{0,1,4\}, \{0,1,6\}, \{0,2,3\}, \{0,2,4\}, \{0,2,5\}, \{0,2,6\},\\
\{0,3,5\}, \{0,4,6\}, \{1,2,4\}, \{1,2,6\}, \{1,3,4\}, \{1,3,5\}, \{1,3,6\},\\
\{1,4,5\}, \{1,5,6\}, \{2,3,5\}, \{2,4,6\}, \{3,4,6\}, \{3,5,6\}, \{4,5,6\}\\
\end{array}
\end{equation}
The 14 remaining marginal density matrices associated with bipartitions
\begin{equation}\label{bipartitions7_aNOOK}
\begin{array}{l}
\{0,1,3\}, \{0,1,5\}, \{0,3,4\}, \{0,3,6\}, \{0,4,5\}, \{0,5,6\}, \{1,2,3\},\\
\{1,2,5\}, \{1,4,6\}, \{2,3,4\}, \{2,3,6\}, \{2,4,5\}, \{2,5,6\}, \{3,4,5\}
\end{array}
\end{equation}
are not completely mixed, having $\mathrm{Tr}(\rho^2)= 0.156250$ in all the cases
except for $\{1,4,6\}$, which is 0.218750.

A different 7-qubit state with high entanglement and very simple algebraic
structure is

\begin{equation}\label{state7_b}
\begin{array}{rl}
|\psi_7^b\rangle = & \frac{1}{4\sqrt{2}}\big(
(1+i)(|0011101\rangle + |0100010\rangle + |0111000\rangle + |1101100\rangle\\
& + |1111011\rangle - |0001001\rangle - |1000111\rangle - |1110101\rangle)\\
& + (1-i)(|0010011\rangle + |0110110\rangle + |1001010\rangle + |1010000\rangle\\
& + |1011110\rangle - |0000100\rangle - |0101111\rangle - |1100001\rangle)
\big)\\
\end{array}
\end{equation}
\begin{equation}\nonumber
E_{\mathrm{NPT}}|\psi_7^b\rangle = 152.504073
\end{equation}

The analysis for $|\psi_t^b\rangle$ is completely similar to the previous case.
Again, all the marginal density matrices for subsystems of 1 and 2 qubits are
completely mixed. In the case of 3-qubit subsystems, the 18 marginal density
matrices corresponding to bipartitions
\begin{equation}\label{bipartitions7_bOK}
\begin{array}{l}
\{0,1,2\}, \{0,1,3\}, \{0,1,4\}, \{0,1,5\}, \{0,2,6\}, \{0,3,5\},\\
\{0,4,6\}, \{1,2,6\}, \{1,3,5\}, \{1,4,6\}, \{2,3,4\}, \{2,3,5\},\\
\{2,3,6\}, \{2,4,5\}, \{2,5,6\}, \{3,4,5\}, \{3,4,6\}, \{4,5,6\}
\end{array}
\end{equation}
are completely mixed, though not the remainder 17 corresponding to bipartitions
\begin{equation}\label{bipartitions7_bNOOK}
\begin{array}{l}
\{0,1,6\}, \{0,2,3\}, \{0,2,4\}, \{0,2,5\}, \{0,3,4\}, \{0,3,6\},\\
\{0,4,5\}, \{0,5,6\}, \{1,2,3\}, \{1,2,4\}, \{1,2,5\}, \{1,3,4\},\\
\{1,3,6\}, \{1,4,5\}, \{1,5,6\}, \{2,4,6\}, \{3,5,6\}
\end{array}
\end{equation}
The squared marginal density matrix have in all cases a trace of 0.156250,
except for the case $\{2,4,6\}$, which is 0.187500.

\subsection{8-qubit states}
The search for 8-qubit states is not as successful as in previous cases. An important
factor in this case is definitely the computational hardness of exploring numerically the space.
Each evaluation of the fitness function on a candidate state takes around 7 seconds.
This constitutes a serious limitation to the amount of states that can be explored in
reasonable time. With the computing power available to us, a typical 4-day search visits
around 40000 states (apart from the fitness evaluation, the search algorithm also
imposes an overhead). In the case of 5, 6, and 7 qubits the search finishes well before
40000 states are visited, but this seems not to be the case here. We empirically found that
a possible reason for this poor performance is related to the size of the neighbourhood.
According to our move function, each candidate state has $2^n(|V|-1)$ neighbours. For a fixed
$V$, the size of the neighbourhood increases exponentially in the number of qubits. In the
case of 8 qubits, the average number of neighbour states visited before a new state is accepted
is considerably higher than in previous cases, resulting in a very slow progression of the search.

According to the NPT measure, the maximum possible (attainable or not) entanglement for a 8-qubit
state is 504.5. We have found states with entanglement up to around 440 and rather simple structure.
As an example, the state
\begin{equation}\label{state8}
\begin{array}{rl}
|\psi_8\rangle = & \frac{1}{8}\big(\\
& |00000010\rangle - |00000101\rangle - |00001111\rangle + |00010111\rangle + |00011010\rangle -\\
& |00011011\rangle - |00100001\rangle - |00100101\rangle + |00100110\rangle + |00101000\rangle +\\
& |00101110\rangle + |00101111\rangle - |00110000\rangle - |00110011\rangle + |00110100\rangle -\\
& |00111100\rangle - |01000000\rangle + |01000110\rangle + |01000111\rangle - |01001010\rangle +\\
& |01001100\rangle - |01001101\rangle + |01010001\rangle + |01010110\rangle + |01011010\rangle +\\
& |01100010\rangle - |01100011\rangle + |01101001\rangle + |01111001\rangle + |01111101\rangle +\\
& |01111110\rangle + |10000001\rangle + |10000111\rangle + |10001000\rangle + |10001010\rangle +\\
& |10001100\rangle - |10001110\rangle + |10010000\rangle + |10010100\rangle + |10011101\rangle +\\
& |10100010\rangle - |10101001\rangle + |10101011\rangle - |10110111\rangle + |10111001\rangle -\\
& |10111010\rangle - |10111111\rangle - |11000011\rangle - |11000101\rangle - |11001011\rangle -\\
& |11010000\rangle + |11010011\rangle + |11010100\rangle - |11011011\rangle + |11011111\rangle +\\
& |11100000\rangle + |11100100\rangle - |11101101\rangle - |11110101\rangle - |11110111\rangle +\\
& |11111000\rangle + |11111010\rangle - |11111100\rangle + |11111110\rangle
\big)\\
\end{array}
\end{equation}
has $E_{\mathrm{NPT}}|\psi_8\rangle = 439.302328$ and only 64 non-null coefficients
equal to $\pm 1$. The 8 single-qubit marginal density matrices are almost completely mixed,
with $\mathrm{Tr}(\rho^2)$ being $0.500488$ (5 times), $0.500977$ (3 times) and $0.5$ (1 times).
Something similar happens for the 28 2-qubit density matrices, the average $\mathrm{Tr}(\rho^2)$
being $0.253540$. The 3- and 4-qubit marginal density matrices deviates considerably more, on
average, from the expected value of a completely mixed state.

\subsection{Performance and state evolution}
For 5 and 6 qubits, most runs result in a maximally entangled state whenever the
parameterisation used for the search algorithm is appropriate. In both cases,
a solution is often found in the first 4-6 cycles (i.e. the search
ends after 4000-6000 generations). This constitutes a fairly rapid convergence,
a typical search taking around 20-30 seconds for 5 qubits and 1-2 minutes for 6 qubits
on a usual laptop. The cases of 4, 7 and 8 qubits are slightly different, as the search never converges
to a maximally entangled state and the procedure ends after a
certain number of consecutive non-improving cycles. A typical
search resulting in a 7-qubit state with $E_{\mathrm{NPT}} \geq 152$ takes around
3-4 hours, while the 8-qubit case requires around 4-5 days. The running time is obviously exponential
in the number of qubits, and the fundamental bottleneck comes from the evaluation of the fitness
function, particularly the entanglement measure.

\begin{figure}[t]
\begin{center}
\includegraphics[scale=0.9]{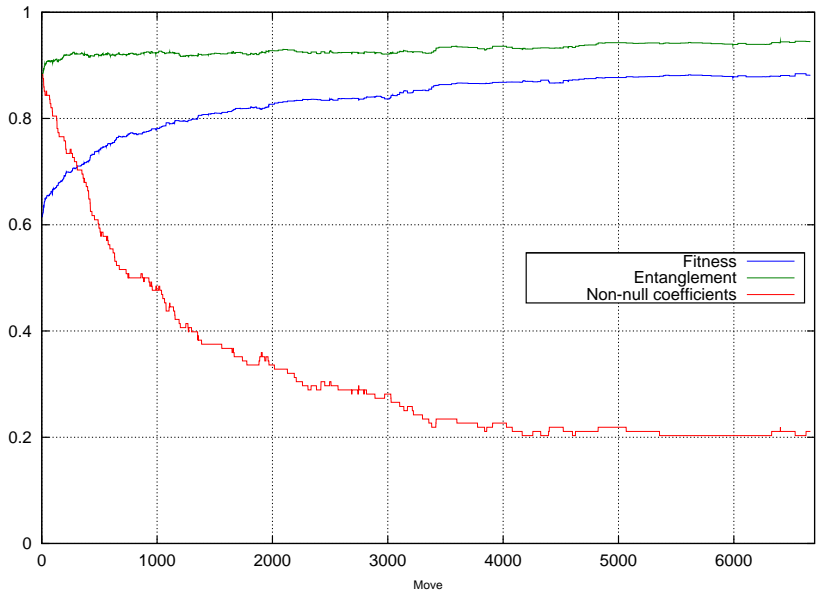}
\end{center}
\caption{Typical state evolution (in colour in the electronic version)}\label{Fig:Evolution}
\end{figure}

Figure \ref{Fig:Evolution} shows the evolution of a typical search for 7 qubits. Apart from the
fitness, both entanglement and the fraction of non-null coefficients are shown. It can be
clearly observed the efect of the punishment factor in the fitness function: after a few
thousands evaluations, candidate states have approximately 80\% of null coefficients. The
overall behaviour is identical for searches of 4, 5, 6, and 8 qubits. (In the case of 5 and 6,
however, the entanglement component eventually reaches the upper bound and the search stops.)

\section{Conclusions and future work}\label{Sec:Summary}
In this work we have shown how states with simple structure can be
discovered by restricting the search to certain subspaces of interest.
For 5 qubits, states equivalent to Brown \etal's $|\psi_5\rangle$ are easily
found. In the case of 6 qubits, a new state with an algebraic structure
arguably simpler than Borras \etal's $|\psi_6\rangle$ state has been discovered.
Similarly to what has been
pointed out for the latter, new teleportation and secret state sharing applications
could be constructed using them. Furthermore, the fact of having at our disposal
\emph{many} of such states might enable us to devise new applications based upon
the use of multiple, non-identical states having maximum entanglement.

For the case of 7 qubits, various states with high entanglement and very simple
algebraic structure have been found. Even though the entanglement is not maximum,
all the single- and 2-qubit marginal density matrices are completely mixed, as
well as more than half of the 3-qubit marginal density matrices. These features
make of them potentially useful to support some applications.

The case of 8 qubits is hard to explore numerically with our available
computational resources. Although the experimentation has not been as intensive
as before (and therefore the parameterisation employed might not be adequate for
this case), our preliminary results seems promising. Compact 8-qubit states
with higher entanglement might be found with more computational resources.

In this work we have restricted the search to \emph{pure} multi-qubit states.
Consequently, the use of the negativity of the partial transpose as measure of
entanglement is a valid but unnecessary choice. An entropic measure such as the von
Neumann entropy should be enough in terms of providing adequate search
guidance.

An interesting avenue for future research is related to a further work of Borras
\etal \cite{Borras09}. Here the authors explore the robustness of entangled states
under various decoherence channels. Along these lines, we plan to study the
entanglement decay of the states introduced in this paper.

\ack
The authors would like to thank Sam Braunstein, Anthony Sudbery, Stephen Brierley, Stefan Weigert
and Rebecca Ronke for helpful comments and stimulating discussions. Feedback from an
anonymous reviewer provided us with valuable suggestions which helped to improve
this work.

\section*{References}


\end{document}